# Temperature Dependence of Extended and Fractional $SU(3)$ Monopole Currents

Ken Yee

Dept. of Physics and Astronomy, L.S.U.

Baton Rouge, Louisiana   70803-4001

email: kyee@rouge.phys.lsu.edu

## Abstract

We examine in pure $SU(3)$ the dependence of extended monopole current $k$ and cross-species extended monopole current $k^{\rm cross}$ on temperature $\tau$, monopole size $L$, and fractional monopole charge $\propto 1/q$. We find that features of both $k$ and $k^{\rm cross}$ are sensitive to $\tau$ for a range of $L$ and $q$. For example, the spatial-temporal asymmetry ratios of both $k$ and $k^{\rm cross}$ are sensitive over a range of $L$ and $q$ to the $SU(3)$ deconfinement transition. The motivation for studying cross, extended, and fractionally charged monopoles in $SU(3)$ is explained.

# 1 APQCD

A challenging problem in QCD is to identify its confinement mechanism and understand how it works. To this end "compact" or lattice QED(CQED), whose action is

$$-S_{CQED} = \sum_P \beta_{CQED} \cos \Theta_P, \quad (1)$$

provides a compelling prototype. Upon a BKT transformation [1], the CQED expectation value of a Wilson loop $W \equiv \exp i(A, J)$ in lattice different forms notation is

$$\langle W \rangle \propto \sum_{\{k|\partial k=0\}} \exp\{-\widetilde{S}\}, \quad (2)$$

$$\widetilde{S} \equiv \frac{1}{2\beta_{CQED}}(J, \Delta^{-1}J) + 2\pi^2 \beta_{CQED}(k, \Delta^{-1}k) - 2\pi i(dk, \Delta^{-1*}E). \quad (3)$$

$J$ and $k$ are, respectively, conserved electric and magnetic monopole current loops. $E$ is the electromagnetic field due to external current $J$: $\partial E = J$. The first and second terms of $\widetilde{S}$ correspond to the electromagnetic interaction energies of $J$ and $k$. The third term is the interaction between the monopole currents and the background electric field $E$ created by $J$. When coupling constant $\beta_{CQED}$ is sufficiently small, the entropy of the sum over monopole loops in (2) dominate over suppression by Boltzmann factor $\exp\{-\widetilde{S}\}$ and monopoles condense. In this phase CQED exhibits the dual Meissner effect: monopoles dynamically rearrange the background electric field $E$ into an effective flux tube [2] which gives the Wilson loop its string tension. In this way, magnetic monopole condensation causes electric confinement in CQED.

An analogous demonstration that monopole condensation is the origin of QCD confinement would be a great achievement. To this end 't Hooft suggested the following idea [3]. Suppose QCD monopoles, like the 't Hooft-Polyakov monopoles of the Georgi-Glashow model [4], carry charges that are magnetic with respect to the $[U(1)]^{N-1}$ Cartan subgroup of color $SU(N)$. Then $SU(N)$ gauge symmetry obscures the magnetic charges and it is necessary to gauge fix at least the $SU(N)/[U(1)]^{N-1}$ symmetry to expose them.

Accordingly, decompose the gauge field into diagonal or neutral($n$) and off-diagonal or charged($ch$) parts $A = A^n + A^{ch}$. Gauge fix to a gauge such as maximal Abelian(MA) gauge [5]

$$D^n_\mu A^{ch}_\mu \equiv \partial_\mu A^{ch}_\mu - ig[A^n_\mu, A^{ch}_\mu] = 0 \qquad (4)$$

which leaves a residual $[U(1)]^{N-1}$ symmetry. In MA gauge the residual $[U(1)]^{N-1}$ symmetry is

$$\Omega_{\text{residual}} = \begin{pmatrix} \exp^{i\omega_1} & & \\ & \ddots & \\ & & \exp^{i\omega_N} \end{pmatrix}, \qquad \sum_{i=1}^N \omega_i = 0. \qquad (5)$$

Under this symmetry the $N$ neutral fields $(A^n)_{ii}$ transform as photon fields whereas the $N(N-1)$ charged fields $(A^{ch})_{ij}$ transform as charged matter fields:

$$(A^n_\mu)_{ii} \to (A^n_\mu)_{ii} + \frac{1}{g}\partial_\mu \omega_i, \quad (A^{ch}_\mu)_{ij} \to (A^{ch}_\mu)_{ij} \exp^{i(\omega_i-\omega_j)} \qquad i,j \in [1,N]. \quad (6)$$

Each $(A^{ch})_{ij}$ carries two different $U(1)$ charges so that the $A^{ch}$ fields induce interactions between the $N$ photon species. Since $\sum_{i=1}^N (A^n_\mu)_{ii}$ is an invariant under (6), an irreducible representation of $[U(1)]^{N-1}$ is

$$\theta^i_\mu \equiv (A^n_\mu)_{ii} - \frac{1}{N}\sum_{i=1}^N (A^n_\mu)_{ii}. \qquad (7)$$

These angles transform under $[U(1)]^{N-1}$ as $\theta^i_\mu \to \theta^i_\mu + \frac{1}{g}\partial_\mu\omega_i$ and obey constraint

$$\sum_{i=1}^N \theta^i_\mu = 0. \qquad (8)$$

The $\theta^i$ fields are compact angular variables since they are constructed from matrix elements of an $SU(N)$ gauge field. As such, they may potentially contain monopole currents just like the compact $U(1)$ field of CQED. If 't Hooft's hypothesis is correct, these would-be monopoles are MA gauge manifestations of nonAbelian gauge field features responsible for QCD confinement.

This procedure where QCD monopoles are identified using only the diagonal $\theta^i$ components of the nonAbelian gauge fields is known as Abelian projection [6]. The quantum dynamics of the $N$ $\theta^i$ fields is known as Abelian projection QCD or APQCD, whose action will be denoted $S_{APQCD}$. Formally, $S_{APQCD}$ is obtained by integrating out the $A^{ch}$ fields in MA gauge from the QCD action: [7]

$$S_{APQCD} \equiv -\log\left\{\int [dA^{ch}] \; \exp(-S_{QCD}) \; \delta[D^n_\mu A^{ch}_\mu]\right\}. \qquad (9)$$

$S_{APQCD}$ is a $[U(1)]^{N-1}$ invariant action depending on the $N$ constrained compact $\theta^i$ fields. Numerically, $S_{APQCD}$ is the action which would generate the importance sampling $\theta^i$ configurations currently made by Abelian projecting importance sampling QCD configurations.

In this paper we will concentrate on the physical case, $N = 3$. A general $U(1) \times U(1)$ action consistent with APQCD symmetries is[1]

$$-S_{APQCD} = \sum_{L=1}^\infty \sum_{P(L)} \left\{\sum_{i=1}^3 F(\Theta^i_{P(L)}) + \sum_{i<j=1}^3 G(\Theta^i_{P(L)}, \Theta^j_{P(L)})\right\} + \cdots \qquad (10)$$

where $P(L)$ refers to the square $L \times L$ plaquettes in the lattice and superscripts $i$ and $j$ refer to the three $U(1)$ gauge fields. "$\cdots$" refers to allowed operators we are neglecting, such as nonsquare Wilson loops, Polyakov loops(important at finite temperature), and nonlocal interactions between

---

[1]The operators in (10) are not all independent. For example, by (8), $\cos\Theta^3_P = \cos(\Theta^1_P + \Theta^2_P)$.

Wilson loops of different shapes and sizes. We have also neglected explicit monopole operators, which are gauge invariant. Gauge, hermitian conjugation of the $SU(3)$ links($U_{x,\mu} \to U^\dagger_{x,\mu}$), and species permutation [8] symmetry require that

- $F(x+2\pi) = F(x)$ and $G(x+2\pi, y) = G(x, y)$;

- $F(-x) = F(x)$ and $G(-x, -y) = G(x, y)$;

- $G(y, x) = G(x, y)$.

Therefore, up to a constant $F$ and $G$ must be of the form

$$F(x) = \sum_{q=1}^{\infty} \beta_q \cos(qx), \qquad (11)$$

$$G(x, y) = \sum_{p=1}^{\infty} \sum_{q=1}^{\infty} \left\{ A_{p,q} \cos(px + qy) + B_{p,q} \cos(px - qy) \right\} \qquad (12)$$

where $p, q$ are integers, $A_{q,p} = A_{p,q}$, and $B_{q,p} = B_{p,q}$.

In (2) only $L = 1$, unit charge $q = 1$ monopoles $k$ are elementary dynamical or "integration" variables because $S_{CQED}$ contains only $L = 1$ plaquettes. $L > 1$ and $q > 1$ monopoles are composite(nonelementary) operators in cQED. Influenced by CQED, the most famous model of monopole confinement, previous studies of APQCD monopoles have been mostly restricted to $L = q = 1$. However, $S_{APQCD}$ may in principle contain a whole range of nonzero $\beta_q(L)$ (and also $A_{p,q}$, $B_{p,q}$) coupling constants. Accordingly, extended $L^3$ [9] and fractionally charged $\propto 1/q$ monopole currents [10] are potentially elementary dynamical variables in APQCD. The $L^3$ and $1/q$ monopole currents would arise from the BKT transformation of the corresponding $\cos q\Theta_{P(L)}$ operators in $S_{APQCD}$. Therefore, in this paper we extend the study of $L = q = 1$ monopoles to $q \geq 1$ and $L \geq 1$ for a wide range of $L$ and $q$ combinations in APQCD.

The possible relevance of extended and fractional monopoles is not just an abstract issue having to do with the form of $S_{APQCD}$; it also has a physical bearing. $L > 1$ monopoles have recently come of interest due to the open possibility that QCD monopoles are not pointlike and may have a nonzero physical core radius [11]. In this case it is reasonable to anticipate that the extended monopole operators of appropriate size $L$ would couple better to the underlying physical objects than the usual $L = 1$ operator. Analogously, a similar argument can be made for $q > 1$ monopoles. There is no *a priori* way to know that $q > 1$ monopoles do not contribute in some quantitative way to the string tension or width of some representations of the QCD flux tube.

Additionally, there is the issue of cross-species monopoles. We have so far been discussing "diagonal" monopoles, magnetic kinks in a *single* $\theta^i$ field. Diagonal monopoles correspond to nonzero $\beta_q$ coupling constants. As defined in Section (3), "cross-species" monopoles are kinks in some linear combination of *two* different $\theta^i$ fields. As illustrated in Ref. [10], such monopoles are elementary dynamical variables when corresponding $A_{p,q}$ and $B_{p,q}$ coefficients of $G(x, y)$ are nonzero. Indeed, we will show that in APQCD cross-species monopoles have nontrivial behavior—in fact the same qualitative dependence on $L$, $q$, and lattice temperature $\tau$ as the diagonal monopoles. Therefore, at least phenomenologically, in APQCD cross species monopoles are *equally as interesting* as their diagonal counterparts.

Sections 2 and 3 present our numerical results. While we will provide some commentary, we do not claim to offer final conclusions or interpretations. Section 2 discusses the correspondence between diagonal $L \geq 1$ and $q \geq 1$ monopoles and operators in $S_{APQCD}$. The nonAbelian gauge dependence of the Abelian projection and the temperature $\tau$ dependence of the diagonal monopoles are described. We find that the diagonal monopole den-sity $\rho$ in APQCD has its minimum at $L = q = 1$. As explained, this result suggests $\beta_{q=1}(L = 1)$ is the dominant coupling in $S_{APQCD}$. We also find that the diagonal monopole current $k$ has subtle but detectable temperature variations between and within the confining and finite temperature phases. These variations are such that the ratio of spatial to temporal monopole densities $\mathcal{R}$ is, as long as $L < 1/\tau$, an *approximately* $L$-independent nontrivial indicator for the $SU(3)$ finite temperature deconfinement transition. For a range of $L$ and $q$, $\mathcal{R} \sim 1$ in the confining phase whereas $\mathcal{R}$ decreases very sensitively with temperature in the finite temperature phase. The redundant behavior of $\mathcal{R}$ at different $L$ and $q$ can have two possible origins: either the higher $L$ and $q$ monopoles are kinematically correlated to the $L = q = 1$ monopoles and there is no independent $L > 1$ and $q > 1$ dynamics going on, or the $L > 1$ and $q > 1$ currents are acting independently and they only appear to be correlated because they share common dynamical rules. At this time it is not clear which of these explanations apply; we suspect it is a combination of both.

Section 3 presents some exploratory results for cross-species monopoles as a function of $L$, $q$, and $\tau$. For a range of $L$, the ratios of spatial to temporal cross-species monopole densities are nontrivial indicators for the $SU(3)$ deconfinement transition. Also, they vary sensitively with temperature in the finite temperature phase.

## 2  Diagonal Monopoles

In Toussaint-Degrand notation [12], define the generalized monopole current

$$k_\mu(L, q) \equiv \frac{1}{2\pi} \sum_{P(L) \in C(L,\mu)} \left\{ \left( q \Theta_{P(L)} \right) \mathrm{mod} 2\pi \right\}, \quad q = 1, 2, \cdots \qquad (13)$$

where $C(L,\mu)$ refers to the $L^3$ cube oriented in direction $\mu$ assuming $D = 3+1$ dimensions. When $q = 1$ and $L > 1$, $k$ reduces to the Type I extended monopoles of Ref. [9]. Integer current $k$ is topologically conserved for all $L$ and $q$ provided one uses an extended derivative when $L > 1$. Because $k_4$ carries magnetic charge $\geq 1/qe$ if the fundamental representation Wilson line carries electric charge $Q = e/2$, we say $q > 1$ monopoles are fractionally charged or "fractional." By Dirac's quantization condition, fractional monopoles of charge $1/qe$ can only interact with electric charge $Q \geq qe/2$ Wilson loops.

Our monopole density is calculated as[2]

$$\rho(L,q) \equiv \langle |k_4(L,q)| \rangle = \sum_{\substack{x \\ \text{configs}}} |k_4(L,q)| \Big/ \sum_{\substack{x \\ \text{configs}}} 1 \ . \tag{14}$$

As previously discussed, only $L = q = 1$ monopoles are elementary dynamical dynamical variables in CQED because $S_{CQED}$, given in Eq. (1) contains only $L = q = 1$ plaquettes. Figure 1A depicts a plot of the monopole density in CQED as a function of $q$ above and below the critical point. As shown, while $\rho(1, q > 1)$ monopoles are also suppressed, $\rho$ is most greatly suppressed at $L = q = 1$. As $q$ becomes greater than 1, $\rho$ converges to

$$\rho_R = \frac{7}{15} = .4\overline{6}, \tag{15}$$

the monopole density when links are completely random [15]. Similarly, as $L$ becomes larger the extended links making up $k(L,q)$—which are superpositions of $L = 1$ links—become more and more disordered. Hence $\rho(L >> 1, q) \to \rho_R$ for all q.

---

[2] The sum over $x$ ranges over all dual lattice sites; the cubes of $k$ are permitted to overlap. Thusly defined, $\sum_x k_\mu(L,q)$ vanishes identically on our periodic lattices. When $L = 1$ our normalization of $\rho$ agrees with Ref. [13]. When $L > 1$ our normalization of $\rho$ disagrees with Ref. [14] by a factor of $L^3$. Also, since our $SU(3)$ results presented below are *fixed* at $\beta = 6.0$ we do not need to convert $\rho$ to physical units for interpretation.

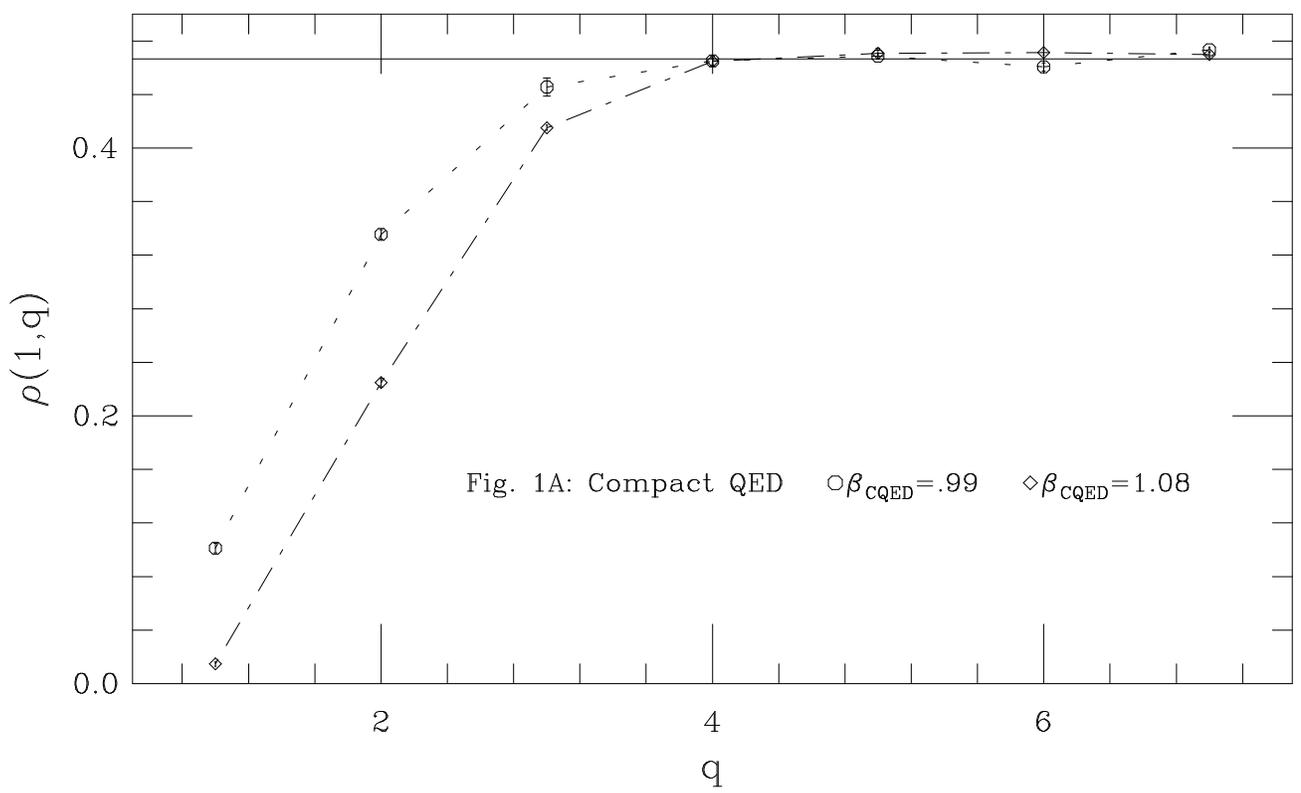

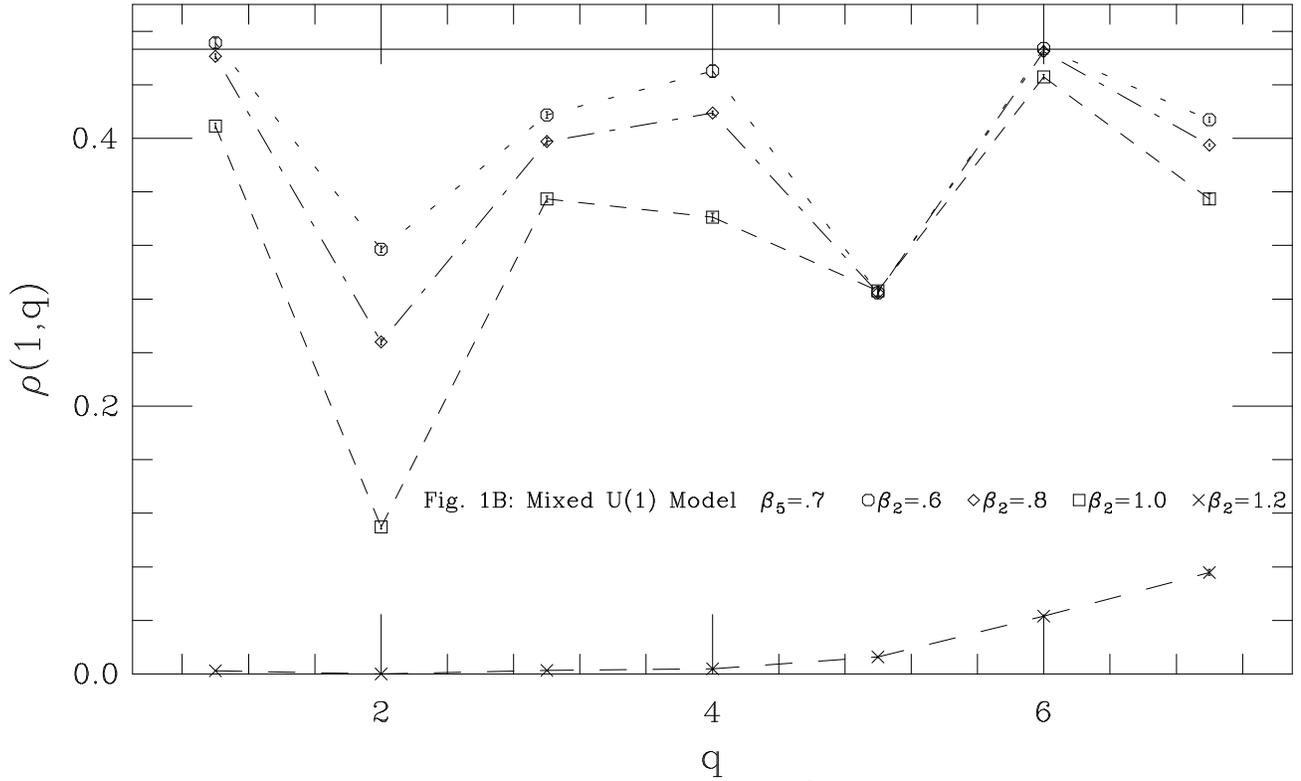

Figure 1: 1A depicts the $q$-dependence of the $1^3$ monopole density in the two phases of CQED; 1B, in a mixed $U(1)$ model with action comprised of $q = 2$ and $q = 5$, $L = 1$ plaquettes. The solid horizontal lines refer to random-link monopole density $\rho_R = 7/15$. The dotted and dashed lines are guides-to-eye. As illustrated, $\rho(L, q)$ equals $\rho_R$ unless it is disrupted by an operator in the action of related $L$ and $q$. $\rho$ is in lattice units.

In a model with $L > 1$ and/or $q > 1$ plaquettes in its action, monopoles of corresponding $L$ and $q$ become elementary dynamical variables. Figure $1B$ depicts $\rho$ as a function of $q$ for the action

$$-S_{mixed} = \sum_{P(1)} \beta_2 \cos(2\Theta_{P(1)}) + \beta_5 \cos(5\Theta_{P(1)}) \qquad (16)$$

at $\beta_5 = .7$ and for a range of $\beta_2$. When $\beta_2 \leq 1.0$, $\rho(1,2)$, $\rho(1,5)$ and to a lesser extent $\rho(1,3)$, $\rho(1,4)$, and $\rho(1,7)$ are suppressed relative to $\rho_R$. When $q >> 5$, $\rho$ uniformly approaches $\rho_R$. On the other hand, at $\beta_2 = 1.2$ the model crosses over to a weak coupling frozen phase where $\rho$ is greatly suppressed for a range of $q$.

In the $\beta_2 \leq 1.0$ phase, much of the $q$-dependence of $\rho$ depicted in Figure 1 can be understood in terms of strong coupling arguments. If the action is zero, the $U(1)$ links are random and $\rho = \rho_R$. If we turn on a representation $q$, $L \times L$ plaquette in the action, then such a plaquette can dress $k(L,q)$ and drive $\rho(L,q)$ below $\rho_R$. If a second plaquette operator is turned on, then $k$ can be dress by the two plaquettes separately and in combination. In Figure 1 $\rho(1,3)$, $\rho(1,4)$, and $\rho(1,7)$ are suppressed because representations $q = 3, 4, 7$ are products of $q = 2$ and $q = 5$. $\rho(1,1)$ and $\rho(1,6)$ are much less suppressed because $q = 1$ and $q = 6$ cannot be made from 2 and 5. As $\beta_2$ increases to 1.2, the strong coupling picture breaks down.

The strong coupling picture predicts the following behavior for the $L$-dependence of $\rho$. If the action contains only plaquettes of size $L_o$ then $\rho(L < L_o, q) \sim \rho_R$ since it is geometrically very difficult for larger plaquettes to dress smaller monopole operators. Similarly, $\rho(L > L_o, q)$ would differ from $\rho_R$ only due to high order strong coupling graphs. Hence only $\rho$ at $L_o$ and $q_o$ associated with corresponding operators in the action (or simple products thereof) would be appreciably suppressed.

All APQCD results presented in this paper are on $24^3 \times T$, $\beta = 6.0$

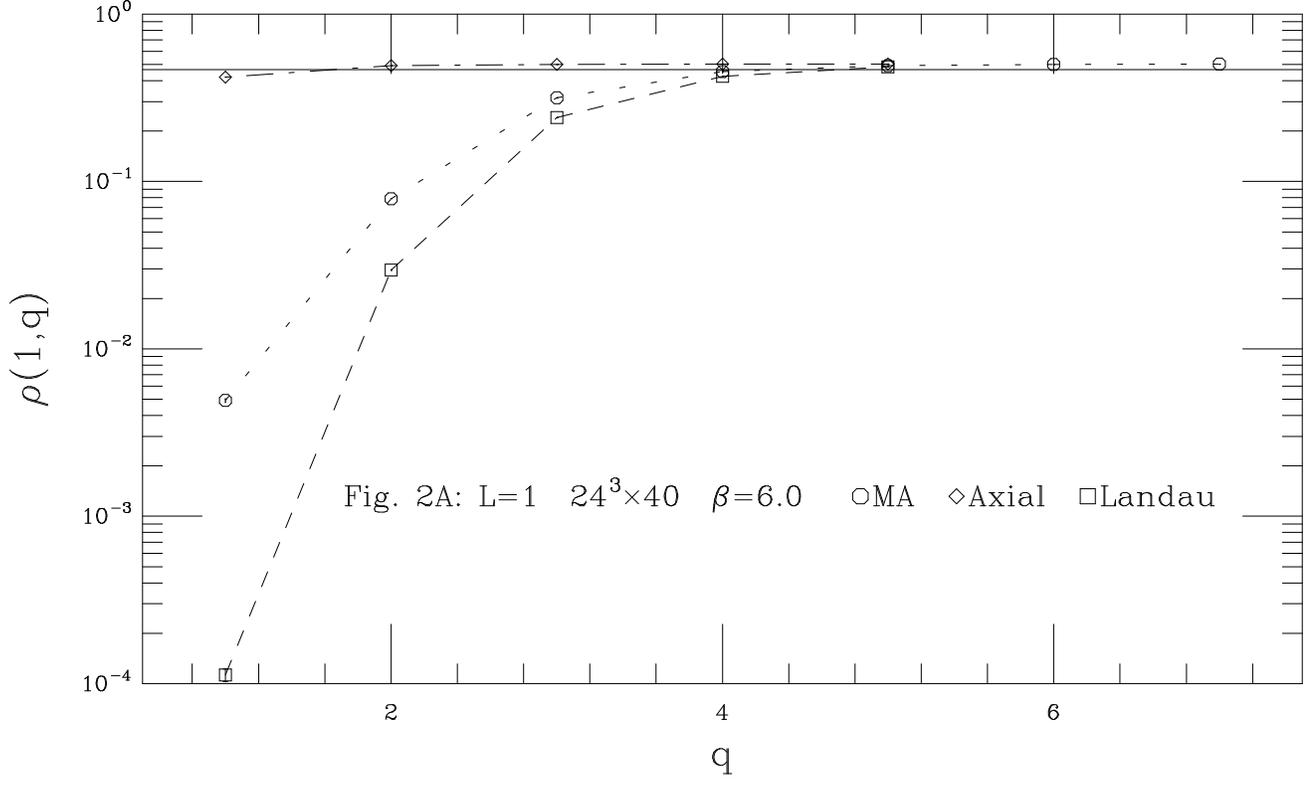

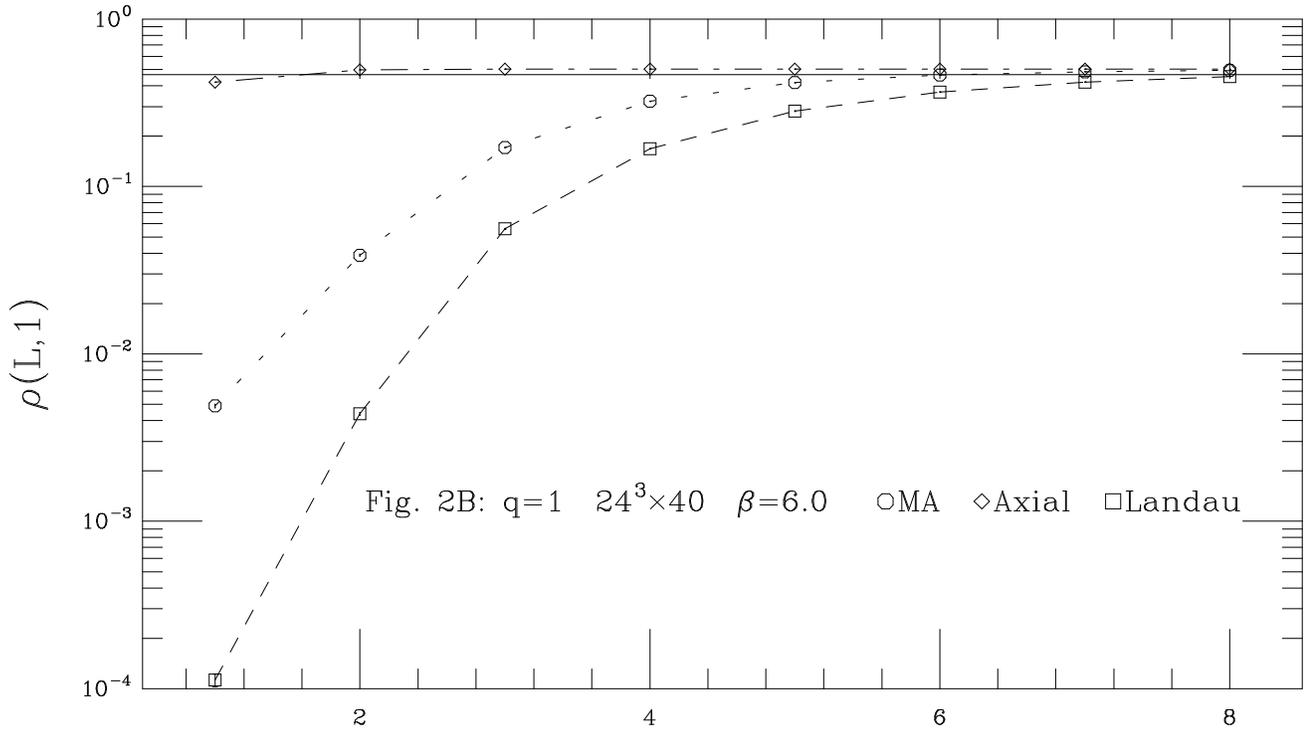

Figure 2: 2A and 2B depict the $q$ and $L$ dependence of the APQCD monopole density $\rho$ in different gauges. $\rho$ is given in dimensionless lattice units. In the three gauges examined, the $q = L = 1$ monopoles are disrupted from the random $\rho_R = 7/15$ value more than $q \neq 1$ or $L \neq 1$ monopoles. Jackknife error bars are, in principle, drawn for all data points in this paper, although sometimes they are too small to be visible.

lattices with periodic boundary conditions. Temperature $\tau$ is given in lattice units as $\tau \equiv 1/T$. On our lattices the $N$ Abelian Polyakov loops $\langle P_i \rangle$ in APQCD vanishes when $T = 40$ and $T = 8$, and is nonzero when $T = 6$ and $T = 4$. Accordingly, the $T = 40$ and $T = 8$ lattices are below and $T = 6$ and $T = 4$ are above the deconfinement temperature $\tau_c$. In the Figures, all quantities are plotted in lattice units with (sometimes unresolvable) jackknife error bars. The dotted lines are guide-to-eye lines; the solid horizontal line indicates the value of $\rho_R$.

In the confining phase of APQCD—corresponding to the confining phase of QCD [6, 16]—one can hope to obtain an indication of the dominant operators in $S_{APQCD}$ from $\rho(L,q)$ in APQCD. Figures 2A and 2B are plots of $\rho$ at $T = 40$ in MA gauge and, for comparison, Landau and axial gauges. The Landau gauge over-fixes the gauge—it leaves no residual $[U(1)]^{N-1}$ symmetry. This seems to "over-suppress" the monopole density relative to MA gauge for a reason we do not fully understand.[3]

In addition to what is shown, we have also evaluated $\rho$ at other $L$ and $q$ combinations and, furthermore, produced similar results on $16^3 \times 24$, $\beta = 5.7$ lattices. In all cases, we find $\rho$ assumes its minimum value[4] at $L = q = 1$, and that $\rho$ is a monotonically increasing function of $L$ for fixed $q$ and $q$ for fixed $L$. Note that while Figure 2A is qualitatively similar to Figure 1A, $\rho(L,q)$ in APQCD is quantitatively very different from its CQED counterpart. For example, in MA gauge $\rho(1,2)/\rho(1,1) = 16.0(.0003)$ whereas this ratio in CQED is $3.32(.004)$ at $\beta_{CQED} = .99$ and $15.3(.003)$ at $\beta_{CQED} = 1.08$. In agreement with an earlier analysis based on plaquette spectral densities [7], this suggests that between $\beta = 5.7 - 6.0$ the biggest coupling in $S_{APQCD}$ is $\beta_{q=1}(L = 1)$. The last statement has two caveats. Firstly, we certainly do not rule out the existence or significance of other $\beta_q(L)$ coupling constants. Secondly, any substantial nonzero $A_{p,q}$ or $B_{p,q}$ coefficients in $S_{APQCD}$ will influence $\rho(L, q)$, possibly obscuring our interpretation.

---

[3]Note that one can always restore the $[U(1)]^{N-1}$ symmetry by making Landau-gauge-violating random $[U(1)]^{N-1}$ gauge transformations. This yields an Abelian projection with the same monopole currents as in Landau gauge because the Landau gauge monopole currents are invariant by construction under $[U(1)]^{N-1}$ gauge transformations.

[4]If we let integer $q$ take on fractional values, the minimum nonzero value of $\rho$ occurs at $q = \frac{1}{2}$ corresponding to magnetic charges $\pm 2$. Since lattice monopole charges don't exceed magnitude 2, $\rho(L, q < \frac{1}{2}) = 0$.

Figures 3 and 4 depict the temperature $\tau$ dependence of $q = 1$ monopoles in MA gauge for a range of $L$. $q > 1$ monopoles(not shown) behave similarly although, since $\rho(L, q > 1)$ is closer than $\rho(L, 1)$ to $\rho_R$, they are less sensitive indicators than $q = 1$ monopoles. While Figure 3 may seem to merely extend the $L = q = 1$ $SU(2)$ results of [17] to $L \geq 1$ in $SU(3)$, let us mention the following technical distinction which is nontrivial. In Ref. [17] the high temperature phase was achieved by varying $\beta_{QCD}$ on a lattice with a fixed number of timeslices. Varying $\beta_{QCD}$ has the disadvantage that it simultaneously changes the physical size of the $1^3$ monopole operator. Thus, this method introduces a hidden variable which may (or may not) have an effect on the observed monopole density. On the other hand, in this paper we fix $\beta_{QCD} = 6.0$ and achieve finite temperatures by varying the number of timeslices $T$. Since our physical lattice spacing is fixed, the physical size of our $1^3$ monopole operator is fixed and we are offering a truly independent verification of Ref. [17]'s original result.

Figures 3A and 3B depict the behavior of the spatial and temporal monopole densities for $L = 1 - 4$. The temporal density is $\rho$ of Eq. (14); the spatial densities are analogously defined using the spatial $k_\mu$ components. The temporal and spatial densities are only mildly $\tau$-dependent with the $L = 1$ densities being the most sensitive. As was noted in Ref. [17] for $L = q = 1$ $SU(2)$ monopoles, contrary to naive expectations *APQCD monopole densities do not decrease dramatically across the deconfinement*

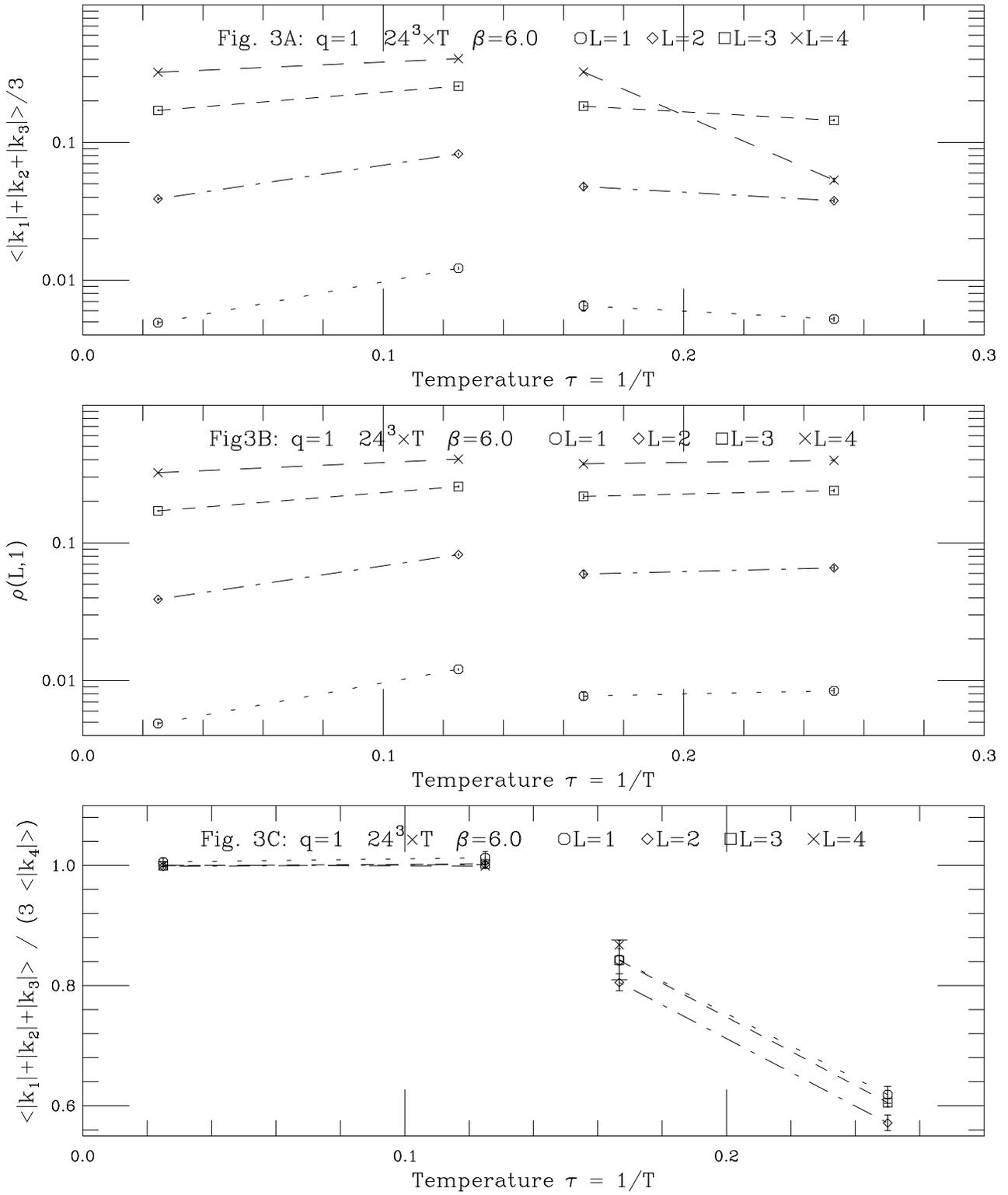

Figure 3: 3A and 3B show that the spatial and temporal monopole densities both decrease across the deconfinement transition and both increase within the confined phase with increasing $\tau$. Within the deconfined phase the spatial density decreases whereas the temporal density rises with increasing $\tau$. Therefore, in 3C the spatial-temporal asymmetry ratio $\mathcal{R}$ has enhanced sensitivity to $\tau$ in the deconfined phase for *all* $L$. The $4^3$ monopoles suffers boundary effects at $\tau = 1/4$.

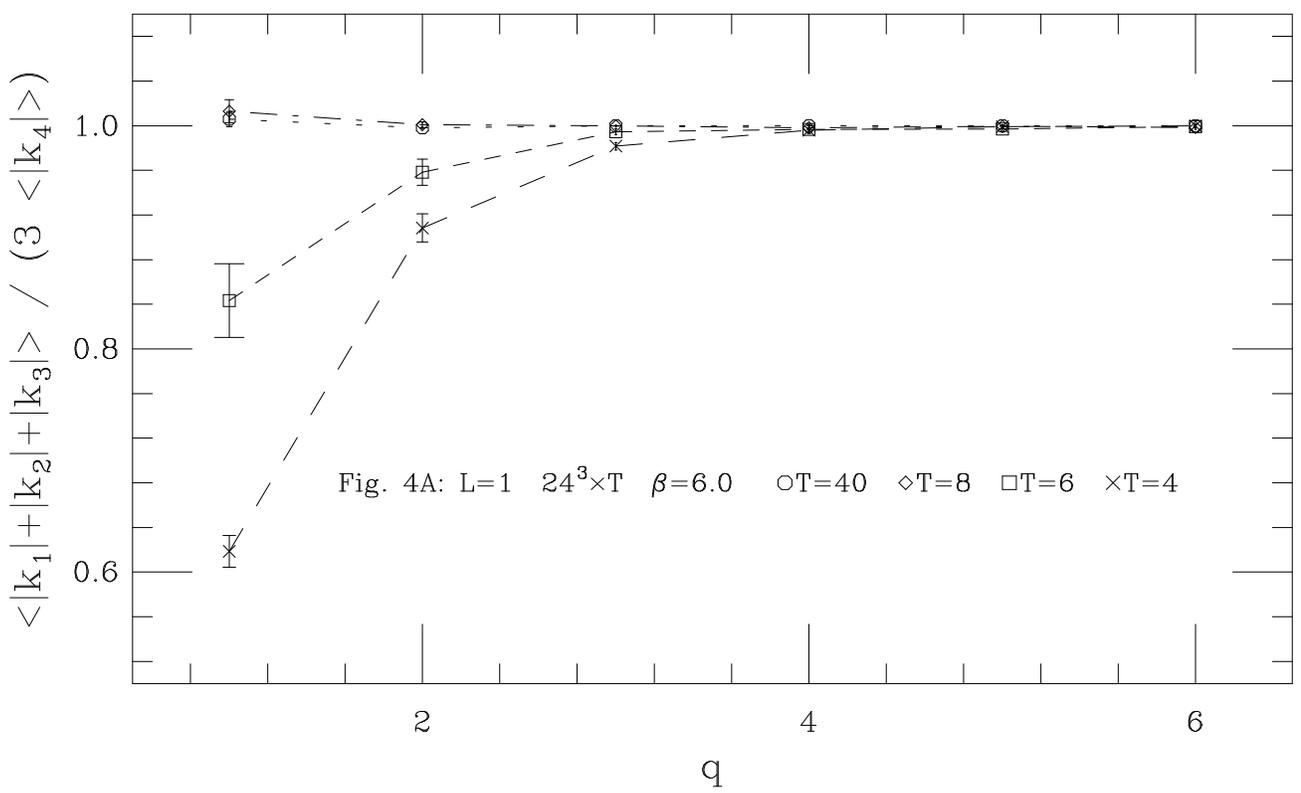

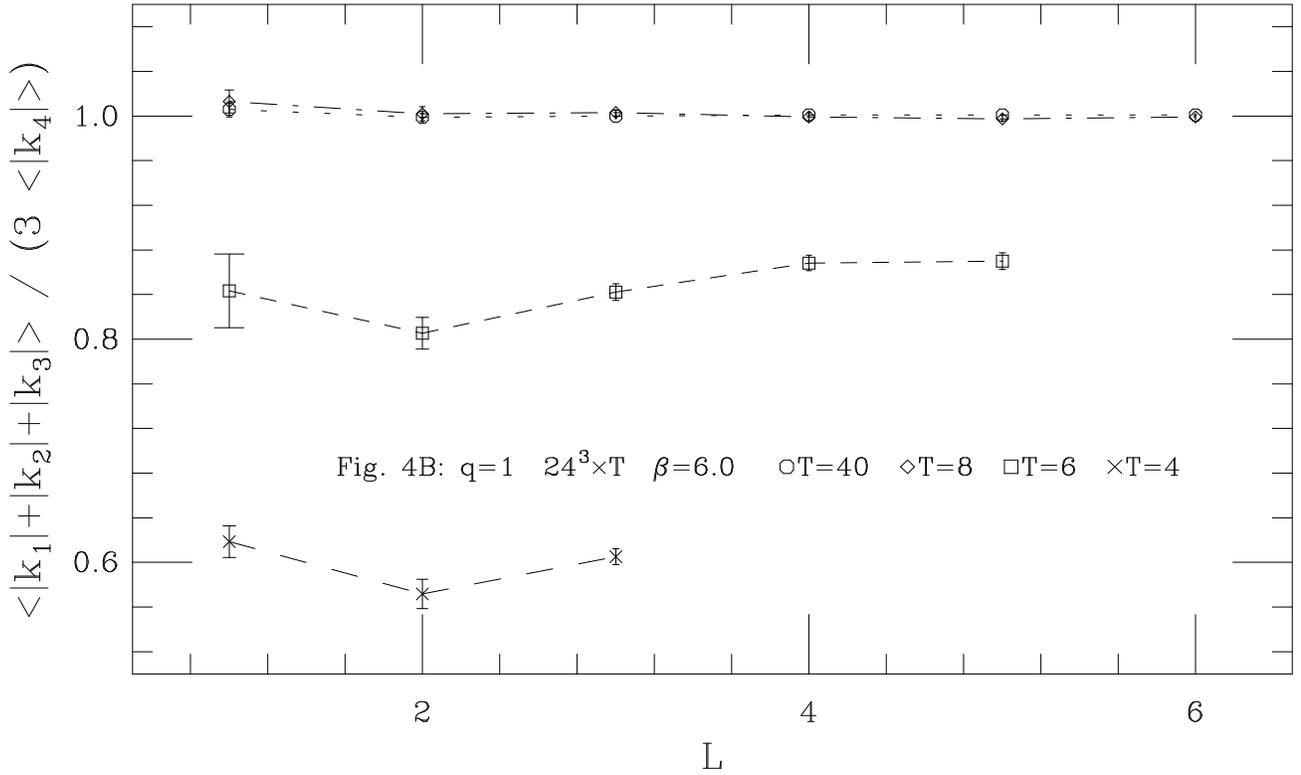

Figure 4: 4A and 4B present the $q$ and $L$ dependence of the ratio $\mathcal{R}$. While $\mathcal{R}$ converges to the uninteresting value of unity at $q > 4$ at all temperatures, it is only very mildly $L$-dependent. At this time we do not understand the significance of such near $L$-independence, which indicates that certain features of $L > 1$ monopoles correlate to temperature as positively as $L = 1$ ones.

*transition* as occurs in CQED. While both densities rise with $\tau$ within the confinement phase, the spatial density falls while the temporal density mildly rises with $\tau$ in the high temperature phase.

There are two comments to be made here. Firstly, monopoles are needed at finite temperature to account for the string tension of *spatially* oriented Wilson loops [18]. Hence, consistency with their confinement-causing role requires that the monopole currents do not disappear at finite temperatures. Secondly, the Abelian Polyakov loop $P_i \equiv \exp i \oint \theta_4^i$ gains a nonzero expectation value at finite temperatures because the time-like link angles freeze (modulo $U(1)$ gauge transformations) to $\theta_4^i \to 0$. Since the spatial components of the monopole currents share time-oriented links with $P_i$, and small link angles tend to decrease the probability of forming a monopole kink, the falloff of spatial densities and the rise of $\langle P_i \rangle$ probably share a common origin: freezing of the time-oriented links with increasing $\tau$. Note that $k_4$ does not contain time-oriented links.

Figure 3C depicts the ratio $\mathcal{R}$ of spatial to temporal monopole densities. As shown, $\mathcal{R}$ is a sensitive finite temperature indicator not just at $L = 1$ [6] but for *all $L = 1 - 4$*.

Figures 4A and 4B reveal the $q$ and $L$ dependence of $\mathcal{R}$ at different temperatures. While $\mathcal{R}$ becomes less and less informative as $q$ increases, what is compelling is that $\mathcal{R}$ *is approximately L-invariant* as long as $L$ is smaller than the smallest lattice width $T$.

# 3  Cross Species Monopoles

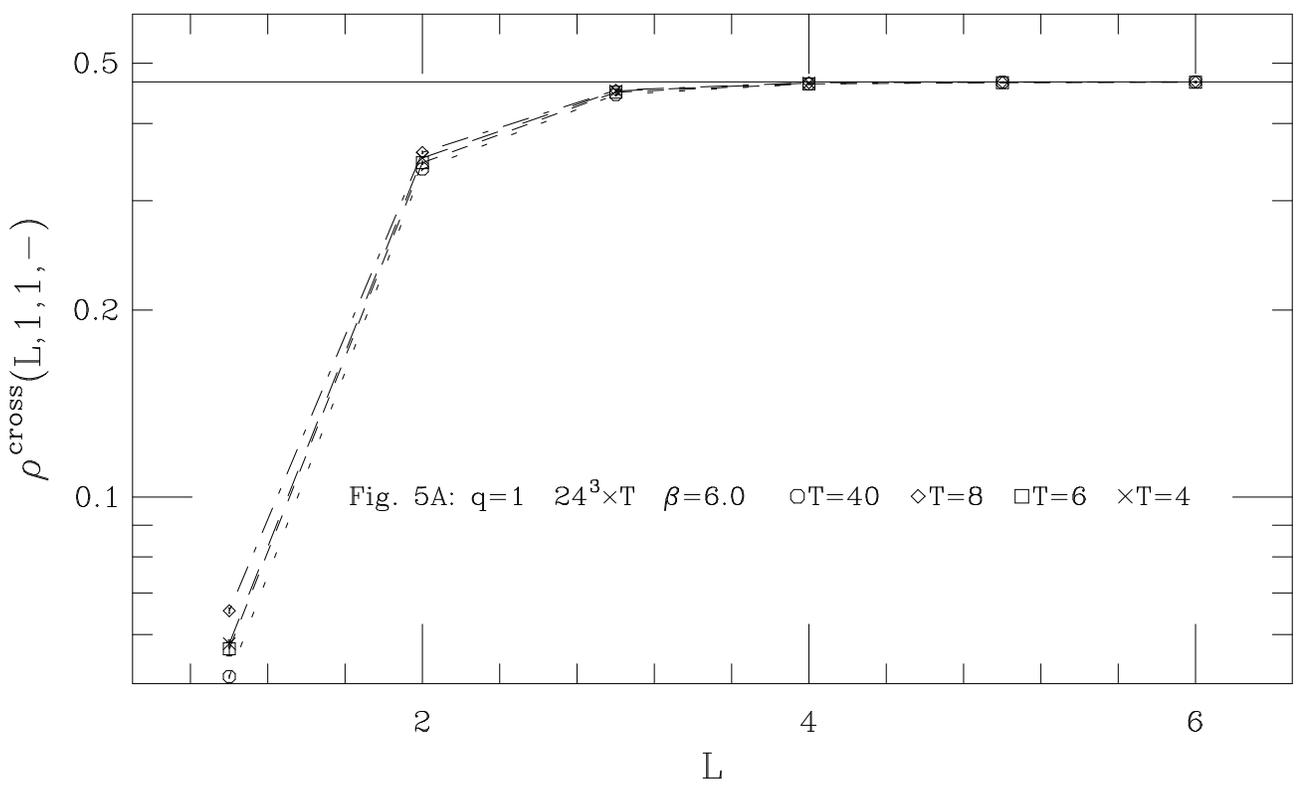

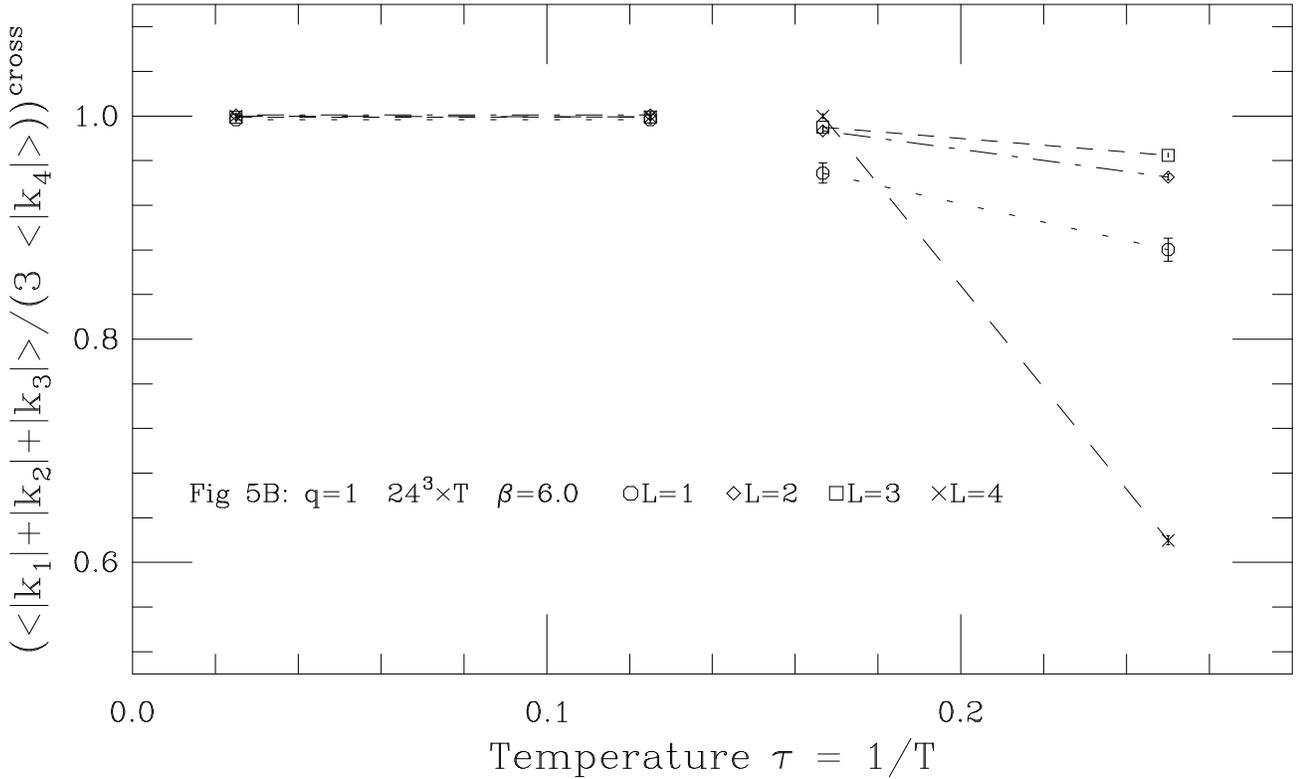

Figure 5: While cross-species monopoles are generally denser than the diagonal ones, they have the same general $q$, $L$, and temperature dependence as the latter. As depicted in 5A the cross-species monopole density is smallest at $L = 1$ and rises quickly to $\rho_R$ at bigger $L$ for all $\tau$. 5B shows that the asymmetry ratio $\mathcal{R}^{\text{cross}}$ is sensitive to temperature in the finite temperature phase. Note the boundary effect on the $L = 4$ current at $\tau = 1/4$.

In addition to $k(L,q)$ one can also consider cross-species monopole currents defined by

$$k_\mu^{\text{cross}}(L,p,q,\pm) \equiv \frac{1}{2\pi} \sum_{P(L)\in C(L,\mu)} \left\{ \left(p\Theta^1_{P(L)} \pm q\Theta^2_{P(L)}\right) \bmod 2\pi \right\} \qquad (17)$$

for integers $p$ and $q$. Cross species monopoles are elementary dynamical variables arising from the interspecies interaction operator $G$ of Eq. (12). For a special case of (10) we illustrated in Ref. [10] how both $k$ and $k^{\text{cross}}$ can simultaneously occur as elementary dynamical variables. In such $U(1)\times U(1)$ models the dual Meissner effect depends on the combined status of $k$ and $k^{\text{cross}}$, which may in principle condense or freeze out *independently* for each value of $L$ and $q$.

In APQCD $k^{\text{cross}}(L,q,q,+)$ is equivalent to $k(L,q)$ by constraint (8). Henceforth we focus on $k^{\text{cross}}(L,q,q,-)$, for which some MA gauge results are presented in Figure 5. $\rho^{\text{cross}}$ and $\mathcal{R}^{\text{cross}}$ are defined analogously to their diagonal counterparts. Our exploratory calculations indicate that generally $k^{\text{cross}}$, while denser, has qualitative features reminiscent of $k$. As demonstrated in Figure 5B, $\mathcal{R}^{\text{cross}}$ is also a nontrivial indicator for the $SU(3)$ finite temperature transition. However, it is noticeably more $L$-dependent than its diagonal counterpart.

We stress that it is unclear at present whether in APQCD $k^{\text{cross}}$ is truly dynamically independent of $k$.

# 4 Acknowledgments

I thank C. Bernard, V. Bornyakov, L. H. Chan, D. Haymaker, Y. Peng, M. Polikarpov, H. Trottier, and R. Woloshyn for their comments. The computing was done at the NERSC(Livermore) Supercomputer Center. The author is supported by DOE grant DE-FG05-91ER40617.